\documentstyle[preprint,epsfig,aps]{revtex} 

\input epsf

\begin{document}

\title{\bf A Phantom Menace? \\
\small Cosmological Consequences of a Dark Energy
Component\\ with Super-Negative Equation of State}

\author{R. R. Caldwell${}^{1}$}

\address{Department of Physics \& Astronomy,
		Dartmouth College,
		Hanover, NH 03755}	
 		
\footnotetext[1]{Robert.R.Caldwell@dartmouth.edu} 

\maketitle  
  
\begin{abstract}

It is extraordinary that a number of observations indicate that we live in a
spatially flat, low matter density Universe, which is currently undergoing a
period of accelerating expansion. The effort to explain this current state has
focused attention on cosmological models in which the dominant component of the
cosmic energy density has negative pressure, with an equation of state $w \ge
-1$.  Remarking that most observations are consistent with models right up to
the $w=-1$ or cosmological constant ($\Lambda$) limit, it is natural to ask
what lies on the other side, at $w<-1$. In this regard, we construct a toy
model of a ``phantom'' energy component which possesses an equation of state
$w<-1$.  Such a component is found to be compatible with most classical tests
of cosmology based on current data, including the recent type 1a SNe data as
well as the cosmic microwave background anisotropy and mass power spectrum.  If
the future observations continue to allow $w<-1$, then barring unanticipated
systematic effects, the dominant component of the cosmic energy density may be
stranger than anything expected.

\end{abstract}

\keywords{cosmology: theory}
 
  
\section*{\null}
\vspace*{-0.5cm}

Arguments have been put forward that we live in a spatially flat, low matter
density Universe which is currently undergoing a period of accelerating
expansion. If the observational evidence upon which these claims are based are
reinforced and strengthened by future experiments, the implications for
cosmology will be incredible. It would then appear that the cosmological fluid
is dominated by some sort of fantastic energy density, which has negative
pressure, and has just begun to play an important role today. No convincing
theory has yet been constructed to explain this state of affairs, although 
cosmological models based on a dark energy component, such as the cosmological
constant ($\Lambda$) or quintessence (Q), are leading candidates. At this
stage we are lead to notice that the parameterization of the dominant energy
component as a fluid  with an equation of state (defined as the ratio of
pressure to energy density) $w(\equiv p/\rho) \ge -1$ leads to the curious
situation that most observational constraints are consistent with models that
go right up to the $w=-1$ border \cite{WCOS99}. It is natural to ask what
lies on the other side of this boundary. The focus of this paper is the
investigation of cosmological models with a dominant component for which
$w<-1$.

We begin by constructing a classical cosmology: a spatially flat FRW space-time
filled by a cold dark matter (CDM) component  and a ``phantom'' (P) energy
component. (A phantom is something which is apparent to the sight or other
senses but has no corporeal existence -- an appropriate description for a form
of energy necessarily described by unorthodox physics, as will be seen later.) 
The phantom energy has positive energy density, $\rho_p>0$, but negative
pressure, such that $\rho_p + p_p < 0$. (It is no fatal flaw for a component to
violate the dominant energy condition for a finite time, as can arise from a
bulk viscous stress due to particle production \cite{Barrow}.) It immediately
follows from the equation of state $w < -1$ that the phantom energy density
grows with time. If this energy has begun to dominate today, it must have come
on like a bolt from the blue. Taking the constant value $w = -3$ as an example,
then $\rho_p \propto a^{6}$ where $a$ is the expansion scale factor. So if
$\Omega_m \sim 0.3$ today, then the Universe contained $90\%$ CDM at $z\sim
0.4$ as opposed to $z\sim 1.8$ in a  $\Lambda$-dominated cosmology. Hence, the
phantom energy exerts its influence very late. We now turn to quantify the
remarkable effect on the cosmology of a component with super-negative equation
of state, $w < -1$.

The expansion scale factor grows rapidly when the phantom energy comes to
dominate the Universe, as $\dot a/a$ actually grows and the deceleration
parameter $q_0 = (1 + 3 w \Omega_p)/2$ becomes very negative. In fact, the
scale factor diverges in finite time: if the expansion is matter dominated
until the time $t_m$,  then  we can write the phantom energy-driven scale
factor as $a(t) = a(t_m)[-w + (1+w)t/t_m]^{2/3(1+w)}$ at $t> t_m$. For $w=-3$
then $a$ diverges when $t=3t_m/2$. Unless $\Omega_m \ll 1$, however, $t_0 < t_m
w/(1+w)$ and so the limiting cosmological time occurs well after the present
day. 

\begin{figure}
\centerline{\null}
\vskip4.0truein
\includegraphics{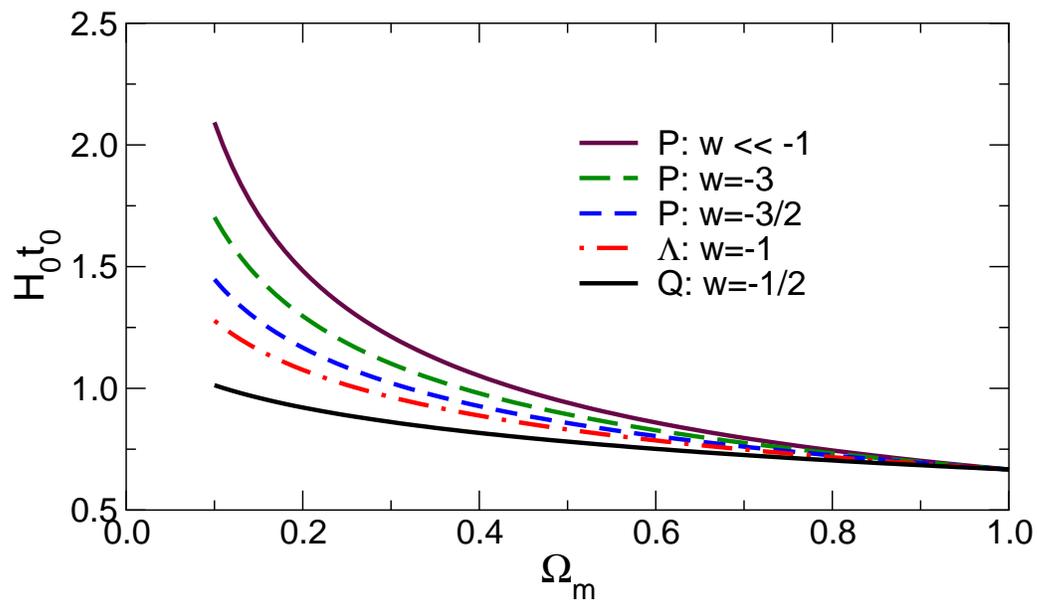}
\caption{The age in units of the Hubble time is plotted versus $\Omega_m$ for a
series of cosmological models containing dark energy with different values of
$w$. \label{figage}} \end{figure}

The expansion age, and similarly the horizon distance in a phantom energy
cosmology are larger than in the analogous $\Lambda$ model. In
Figure~\ref{figage} we show the age for a sequence of $w\le -1$ models. In the
range of interest, near $\Omega_m \sim 0.3 - 0.4$, we see that the phantom
energy can increase the age by up to $\sim 30\%$ over the $w=-1$ age.  For a
given value of $\Omega_m$, as $w$ becomes very negative the phantom energy
becomes important later and later with diminishing effect on the age. In fact,
for $\Omega_m=0.3$, the bound on the age in units of the Hubble time is $H_0
t_0 \le 1.2$.

\begin{figure}
\centerline{\null}
\vskip4.0truein
\includegraphics{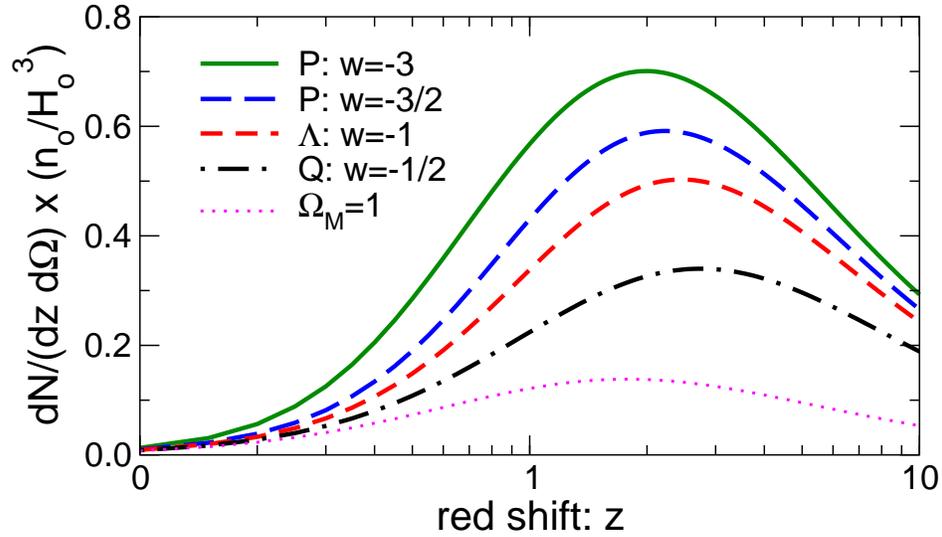}
\caption{The volume - red shift relationship is shown for phantom energy models
with $w = -3,\, -3/2$, $\Lambda$CDM with $w=-1$, QCDM with $w =-1/2$, and CDM.
All the dark energy models have $\Omega_m = 0.3$.  \label{figvol}} \end{figure}

The volume - red shift relationship is demonstrated in Figure~\ref{figvol}. For
the same matter density, the phantom energy model gives the largest
differential number of objects per red shift interval. Although evolutionary
effects are important in cosmological tests based on this relationship, if all
other features are held fixed, the phantom model will predict more strong
gravitationally lensed quasars than $Q$ or $\Lambda$ models.

\begin{figure}
\centerline{\null}
\vskip4.0truein
\includegraphics{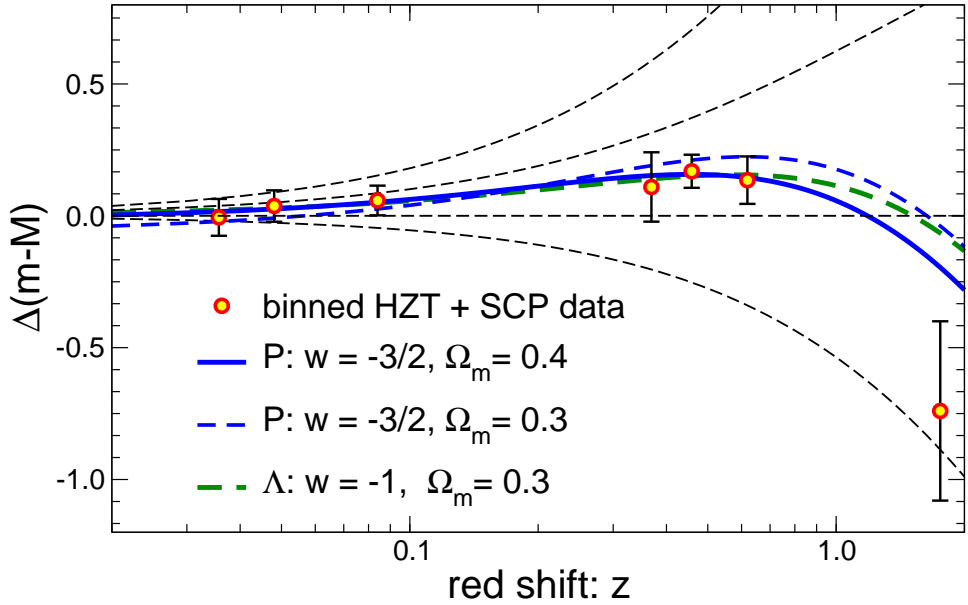}
\caption{The magnitude - red shift relationship is shown for the redshift-binned
type 1a SNe data (HZT--  \protect\cite{mlcsref}; SCP-- \protect\cite{scpref};
binning-- \protect\cite{Riess2001}),  alongside the predictions of various
cosmological models.  Both the phantoms ($w=-3/2, \Omega_m=0.4$; $-3/2,\, 0.3$)
and the $\Lambda$ model ($-1,\,0.3$) provide good fits to the data (low
$\chi^2$/d.o.f.). The magnitudes are calculated relative to an empty, open
Universe. The light, dashed lines are for pure phantom, deSitter, Milne, and
Einstein-deSitter, from top to bottom. \label{figmag} } \end{figure}

The magnitude - red shift relationship is demonstrated in Figure~\ref{figmag}.
The predictions for several cosmological models have been shown along with  a
summary of the observational results. All other parameters being equal, one
expects high red shift supernovae to be dimmer in a phantom energy Universe.  

\begin{figure}
\centerline{\null}
\vskip4.0truein
\includegraphics{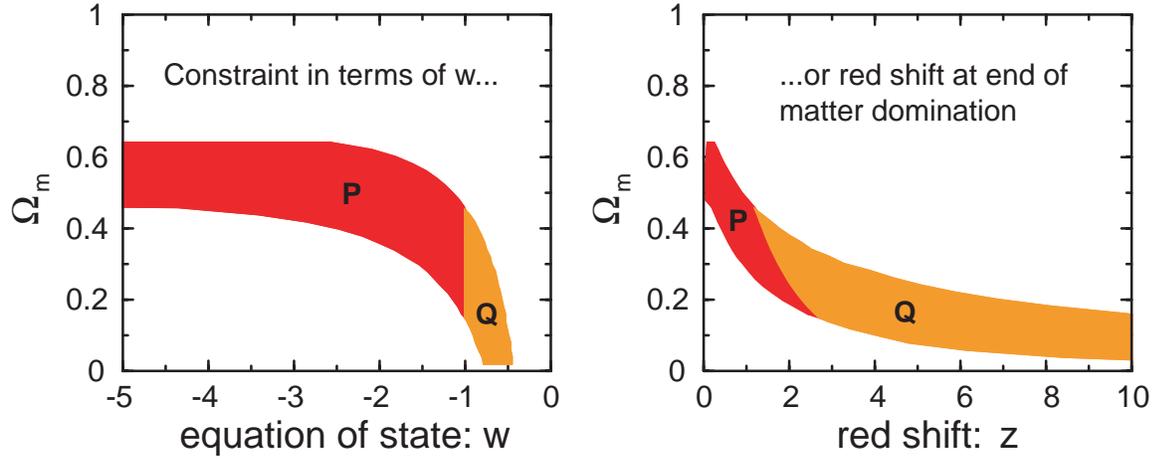}
\caption{The contraint on the allowed values of $\Omega_m$ in phantom and
quintessence dark energy models  is shown as a function of $w$, or 
alternatively the red shift at which matter-domination ends, when $\Omega_m =
0.9$. We have traversed the $w=-1$ boundary, finding that there are phantom
energy models in accord with the observations. \label{figparams}} \end{figure}

Considering the constraint to the allowed range of $w$ and $\Omega_m$ due to the
supernovae, we find that phantom energy models even with very negative $w$ are
in accord with the observations. We see in Figure~\ref{figparams} that the
contour region is extensive for $w<-1$, prefering a slightly higher matter
density than in Quintessence models. An alternative way to visualize this
constraint on phantom models is to construct a parameter space in terms of the
red shift at which the energy density drops to $90\%$ matter, signalling the end
of the matter-dominated epoch, rather than the constant equation of state $w$.
The right-hand panel above makes clear that an important difference with other
dark energy models is the very late onset of the phantom energy.

The next challenge is to construct a microphysical model of the phantom energy
in order to consistently determine the impact on the cold dark matter, baryon,
and photon fluctuation spectra.

We describe the system of gravitation, normal matter, and phantom energy with
the Lagrangian $L = -{R/ 16 \pi G} + L_m + L_p$ where $L_m$ represents the
Lagrangian for the ``normal" matter, CDM plus the standard model particles and
fields. Since it is not possible to achieve $w <-1$ with a free scalar field,
and we wish to avoid interactions with other fields in order to keep the energy
``dark," we start with the unorthodox Lagrangian $L_p = - \partial_\mu \phi
\partial^\mu \phi/2 - V(\phi)$ (with metric signature $+---$). The important
point is that we have switched the sign of the kinetic term in the scalar field
Lagrangian $L_p$. In this way, $\rho_p = -\dot\phi^2/2 + V$ and $p_p =
-\dot\phi^2/2 - V$ so that $w \le -1$ is attained. The equations of motion are
$\ddot \phi + 3 H \dot\phi = +V_{,\phi}$, so that the field will tend to run
up, not down, a potential towards larger energy.  The key ingredient for this
crude toy model, and the more detailed model we introduce later, is the
non-canonical kinetic energy term, examples of which occur in supergravities
\cite{Nilles} and in higher derivative theories of gravity \cite{Pollock}. 

The cosmological spectrum of fluctuations in the phantom field develop in a
fashion similar to the case of Quintessence \cite{CDS97}. The Fourier transform of the
linearized perturbation equation is $\ddot{\delta\phi} + 3 H\dot{\delta\phi} +
(k^2 - V_{,\phi\phi})\delta\phi = - \dot h \dot \phi/2$, where $h$ is the
synchronous gauge metric perturbation. In this equation, the sign of
$V_{,\phi\phi}$ is  different than in the standard case. Just as for
Quintessence, perturbations on small scales, for $k^2 \gg |V_{\,\phi\phi}|$,
are suppressed. On larger scales, however, the phantom energy develops
inhomogeneities in response to the surrounding matter and density
perturbations. If the effective mass, $(k^2 - V_{,\phi\phi})^{1/2}$, should
become imaginary then  $\delta\phi$ would develop a growing (tachyon)
instability. However, if we implement a constant equation of state, then 
$V_{,\phi\phi} = {3 \over 2}(1-w)[\dot H - {3 \over 2}H^2(1+w)]$, which is
negative for a limited range of values of $w$ and $\Omega_m$. Consequently, 
the effective mass is real, and there is no instability for constant $w$.

A more versatile phantom model can be constructed using a Lagrangian with a
non-standard dependence on the field gradients, $L_p = a(\phi) (\nabla\phi)^2
+ b(\phi) (\nabla\phi)^4 + ...$, motivated by the appearance of higher
derivatives in string theory. This approach has been employed in the 
k-inflation and k-essence scenarios \cite{AP99,AP00,AP01}, and it has been
demonstrated \cite{AP99,Chiba00} that such models can also be used to
describe an energy component with a super-negative equation of state, $w_p <
-1$, which is also stable to perturbations (no tachyonic instability). 
Details of the treatment of cosmological perturbations in the k-essence
scenario are given in \cite{Erickson02}; the approach is sufficiently
general that the case of a phantom is simply accommodated. Hence, starting
from a theory giving the functional dependence of $L_p$ on $\phi$ and
$(\nabla\phi)^2$, one can readily determine the stress-energy 
$\rho(z),\,p(z)$ and the sound speed of perturbations $c_p(z)$ as they evolve
in time or red shift. 

Turning this around, one may instead dictate the histories $w_p(z)$ and
$c_p(z)$ in order to specify the phantom dark energy model. In the present
investigation we focus on constant-$w_p$ and $c_p=1$ models. Although this is
not a generic prediction of the dynamical models described above, the
observational differences are not expected to be very large (as some
exploratory analyses have confirmed).  Since the observational effects rely on
a substantial fractional phantom energy density, and since $\rho(z)$ is
actually growing with time, then the relevant values of $w_p(z)$ and $c_p(z)$
are the mean values between the redshift when the phantom began to dominate and
the present. For $w_p < -1$, this is a narrow range in redshift, leaving little
opportunity for the effects of time-variation to be distinct. As we have
modified CMBfast \cite{cmbfast} to include a phantom dark energy component
based on both approaches, we now proceed to discuss the resulting perturbation
spectra.

\begin{figure}
\centerline{\null}
\vskip5.0truein
\includegraphics{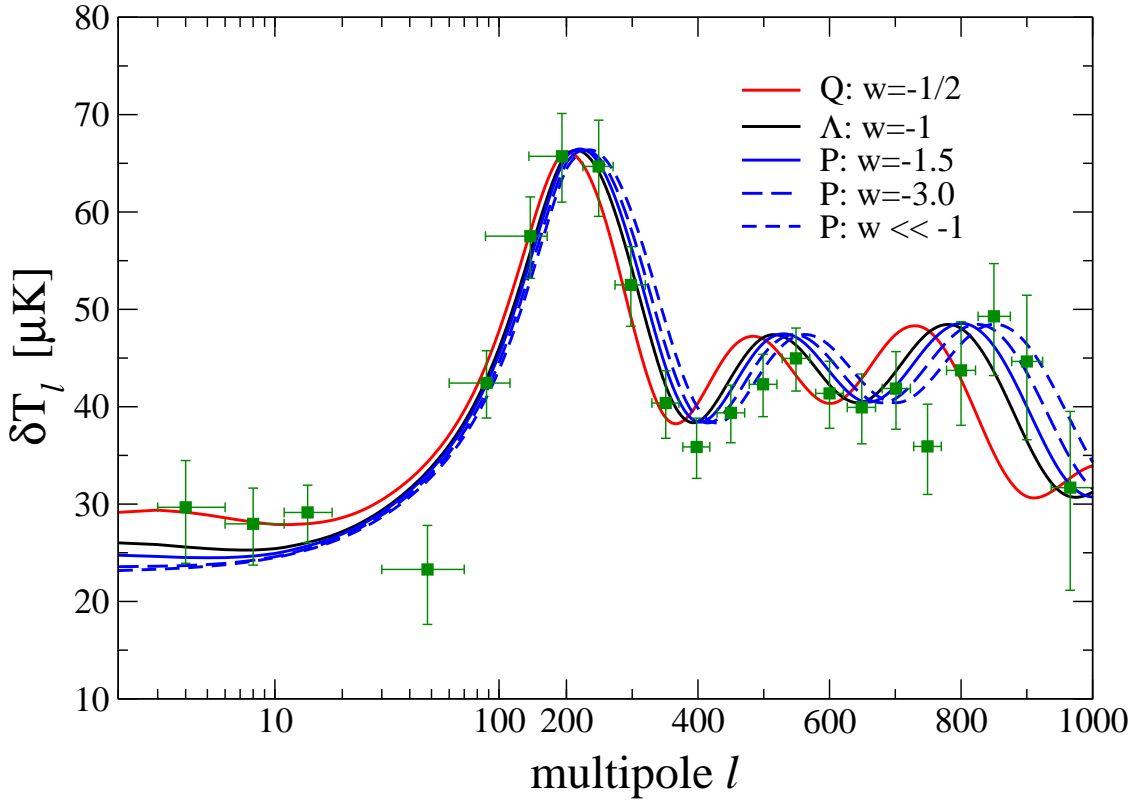}
\caption{The CMB anisotropy spectra are shown for various cosmological models
with equal input power, to demonstrate the effect of $w< -1$. On large angular
scales, the strength of the late-time ISW effect diminishes as $w$ becomes more
and more negative, because the expansion is CDM-dominated until later and later.
The locations of the acoustic peaks shift to smaller angular scales as $w$
becomes more negative, due to the increased distance to the last scattering
surface. The models shown all have $\Omega_m = 0.35$, $\Omega_b h^2 = 0.02$, and
$h=0.70$.  The horizontal axis is logarithmic for $2 < l < 200$  in order to
give enough space to both large and small angular scale features. Comparing the
curves with a compilation of current CMB data  \protect\cite{Tegmark2001}
suggest a more negative dark energy equation-of-state may allow for a better fit
to the small angular scale data. \label{figcmb}} \end{figure}

The main feature that distinguishes the phantom energy case from $\Lambda$ or Q
is that the onset of phantom energy dominance happens at the very last moment
-- so late that most evolutionary effects which occur in $\Lambda$ and Q models
are suppressed. In the cases illustrated in Figure~\ref{figcmb} the strength of
the late-time ISW effect diminishes as $w$ becomes more and more negative,
because the expansion is CDM-dominated until later and later.  On small angular
scales  the locations of the acoustic peaks shift to higher multipole moments
as $w$ becomes more negative, due to the increased distance to the last
scattering surface. This may be an important factor in  attempts to develop  a
phantom dark energy cosmological model to accommodate the newer CMB results,
also represented in Figure~\ref{figcmb}. A more detailed analysis will be
forthcoming. 

\begin{figure}
\centerline{\null}
\vskip4.0truein
\includegraphics{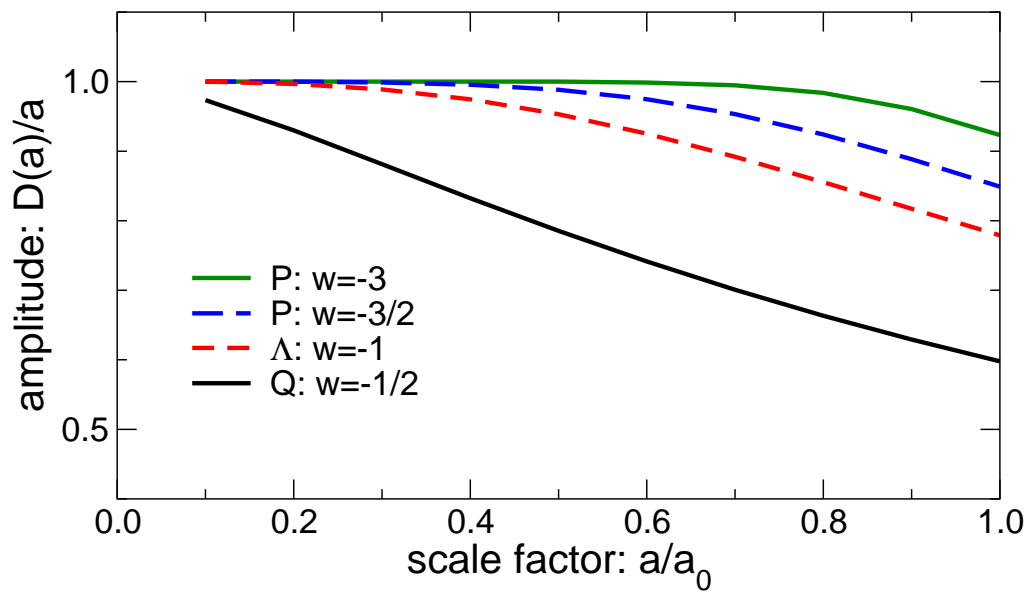}
\caption{The amplitude of the growth suppression factor $g=D(a)/a$ versus the
scale factor is shown for various cosmological models. We see that the
perturbation growth is slowed later, with weaker effect, in the phantom energy
models. The models shown all have $\Omega_m = 0.3$. For SCDM,
$D(a)/a=1$.\label{figgfact}} \end{figure}

The phantom-dominated background evolution has an important effect on the
fluctuation spectra, as well. Making the assumptions  that the matter
component of the cosmological fluid carries a spectrum of scale-invariant
perturbations generated by inflation, and that, similar to Quintessence, the
phantom energy itself does not fluctuate on scales well below the Hubble
horizon, then we may follow the growth of linear perturbations in the CDM and
baryons. The growth suppression factor $g=D(a)/a$ is shown in
Figure~\ref{figgfact}, where we see that perturbations grow as $D \propto a$
until very late,  owing to the very late time at which the phantom energy
begins to dominate. Hence, the evolution is very similar to the standard CDM
scenario, but with a lower matter density.

Next, we consider the behavior of the mass power spectrum for $w<-1$.  Because
the phantom field does not fluctuate on scales well below the horizon, the
overall shape of the mass power spectrum is well-described by the BBKS
parameterization \cite{BBKS}. (See \cite{MaETAL} for instructions on how to
adapt the fitting functions to quintessence). The normalization is obtained by
comparison with the CMB. Using the amplitude of the first acoustic peak, then
the normalization is similar to a $\Lambda$CDM model with the same
cosmological parameters, but with a $\sigma_8$ higher by a ratio of the growth
suppression factors $g_{P}/g_{\Lambda}$. Since $g_{\Lambda}(\Omega_m=0.35,\,
a=1) = 0.8$, and $g_{P}$ is at most unity in the limit of $w \ll -1$, then
$\sigma_8$ is higher by up to $20\%$. Using COBE for the normalization, on the
other hand, the effect of the late ISW and the direct fluctuations in the
phantom field must also be taken into account. However, owing to the late
onset of the phantom, there is very little late ISW and the direct
fluctuations are also negligible. As mentioned earlier, the low-$\ell$
spectrum is very flat, as in SCDM, and up to $\sim 5\%$ lower than in the
comparable $\Lambda$CDM model. Hence, the COBE normalized spectrum will result
in a power spectrum with a $\sigma_8$ up to $\sim 20 + 5\%$ higher than for
$\Lambda$.

\begin{figure}
\centerline{\null}
\vskip5.0truein
\includegraphics{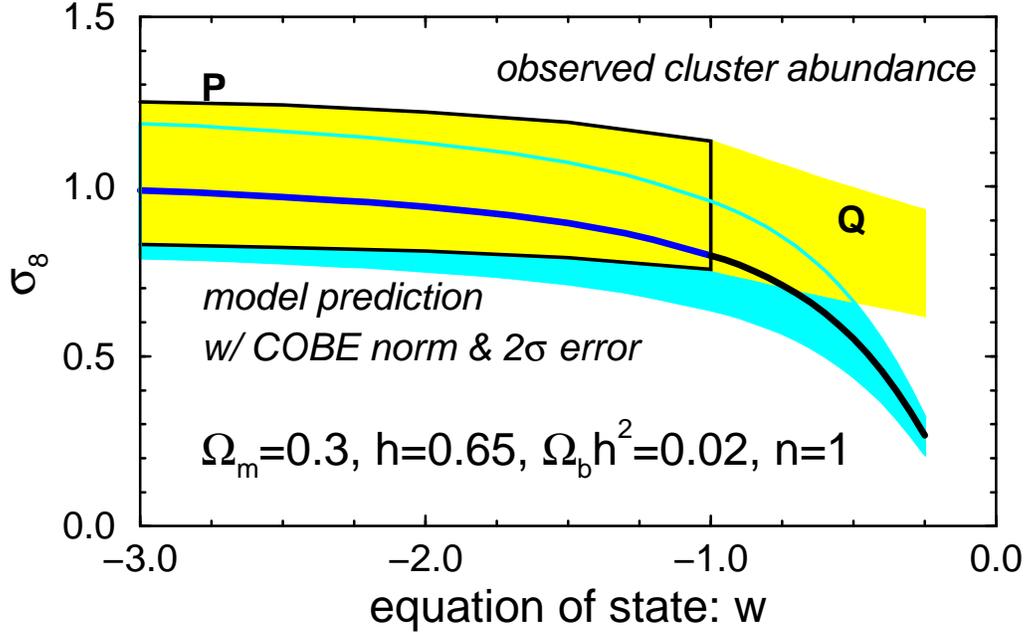}
\caption{The cluster abundance constraint is shown for  a sequence of Q
($w>-1$), $\Lambda$ ($w=-1$), and phantom ($w < -1$) models with
$\Omega_m=0.3,\,h=0.65,\,\Omega_b h^2=0.02,\, n=1$.  The upper shaded region
shows the $2\sigma$ range of $\sigma_8$ based on the observed abundance of
x-ray clusters, as interpreted for a range of $w$ (based on an extension of the
work in Ref. \protect\cite{Wang98}). The lower shaded region shows the
predicted $\sigma_8$ for COBE-normalized models. The intersection of the two
regions grows as $w$ becomes more negative. The letters P and Q indicate the
Phantom and Quintessence  regions, to the left and right of $w=-1$.
\label{figsigma}}
\end{figure}

In Figure~\ref{figsigma} we show the prediction and observational constraint on
$\sigma_8$ for a sequence of models varying in $w$. While the amplitude of the
CMB-normalized mass power spectrum grows slightly as $w$ becomes more negative,
the implied $\sigma_8$ based on the observed abundance of clusters grows as
well. (The cluster abundance calculation is based on an extension of
\cite{Wang98} to the case of $w < -1$.) We see that the intersection of the
$2\sigma$ regions of $\sigma_8$ grows with decreasing $w$. Further constraints
based on large scale structure one may consider, but which are beyond the scope
of this paper, include the evolution of the abundance of rich clusters and the
growth rate of the linear fluctuation spectrum.  Our lesson here is that
phantom dark energy models do exist which, upon first inspection, satisfy
constraints based on the CMB and the mass power spectrum.

In summary, we have investigated the properties of cosmological models in which
the dominant energy density component today has an equation of state $w< -1$.
We have demonstrated the impact on the cosmological age, the volume - red shift
and magnitude - red shift relations, the CMB, and the mass power spectrum,
finding broad agreement with current observational constraints. Current data
appears to be consistent with a phantom, although a  more careful analysis will
be necessary to quantify the observational status of $w<-1$.

While there is no single, best method for determining if the equation of state
is strongly negative, combinations of measurements of the type described in
this paper will substantially improve our understanding of the dark energy. One
of our central points is the importance that such analyses are open to the
possibility of a phantom dark energy -- unjustified biases and priors can lead
to a gross misinterpretation of the observational evidence. (See \cite{Maor01}
for an analysis of the pitfalls of assuming a constant $w$ or $w \ge -1$.)

Of course, we do not want to overlook the distinct possibility that the
observational evidence in favor of such a strongly negative equation of state
is simply a phantom -- that the apparent accelerating expansion is due to more
conventional, though unanticipated causes ({\it e.g.} dust or evolution for the
SNe). If these systematic effects can be eliminated, and the data continue to
support $w<-1$ then the implications for fundamental physics would be
astounding, since $w<-1$ cannot be achieved with Einstein gravity and a
canonical Lagrangian. It is premature to shift attention towards building
$w<-1$ models, but it is important to be aware of the properties and
implications of models in each direction of the cosmological parameter space. 

As has been discussed elsewhere, current observations suggest  the presence of
a dark energy component with $w \lesssim -0.5$ \cite{WCOS99}, with
many constraints pushing towards $w=-1$: the observations are teetering at the
edge of a previously unfamiliar boundary.

\acknowledgements 
The SNe data used in Fig~\ref{figmag} were obtained from Adam Riess \cite{Riess2002}. The cluster
abundance constraint in Figure~\ref{figsigma} was evaluated by Limin Wang. This work was supported by
the US DoE grant DE-FG02-91ER40671 (at Princeton) and the NSF grant PHY-0099543 (at Dartmouth).



\eject

\eject

\eject

\eject

\eject

\eject

\end{document}